\documentclass[twocolumn,showpacs,amssymb]{revtex4}
\usepackage{graphicx}
\usepackage{dcolumn}
\usepackage{bm}
\usepackage{epsfig}

\begin{document}

\title{Testing Einstein's Equivalence Principle with Fast Radio Bursts}

\author{Jun-Jie Wei$^{1}$, He Gao$^{2}$,
Xue-Feng Wu$^{1,3\ast}$, and Peter M{\'e}sz{\'a}ros$^{4,5,6}$}

\affiliation{$^1$ Purple Mountain Observatory, Chinese Academy of Sciences, Nanjing 210008, China \\
$^2$ Department of Astronomy, Beijing Normal University, Beijing 100875, China\\
$^3$ Joint Center for Particle, Nuclear Physics and Cosmology, Nanjing
University-Purple Mountain Observatory, Nanjing 210008, China\\
$^4$ Department of Astronomy and Astrophysics, Pennsylvania State University, 525 Davey Laboratory, University Park, PA 16802\\
$^5$ Department of Physics, Pennsylvania State University, 104 Davey Laboratory, University Park, PA 16802\\
$^6$ Center for Particle and Gravitational Astrophysics, Institute for Gravitation and the Cosmos, Pennsylvania State University, 525 Davey
Laboratory, University Park, PA 16802\\
$^\ast$Electronic address: xfwu@pmo.ac.cn}

\date{\today}


\begin{abstract}
The accuracy of Einstein's Equivalence Principle (EEP) can be tested with
the observed time delays between correlated particles or photons that are emitted
from astronomical sources. Assuming as a lower limit that the time delays
are caused mainly by the gravitational potential of the Milky Way,
we prove that fast radio bursts (FRBs) of cosmological origin can be used to
constrain the EEP with high accuracy. Taking FRB 110220 and two possible
FRB/gamma-ray burst (GRB) association systems (FRB/GRB 101011A and FRB/GRB 100704A)
as examples, we obtain a strict upper limit on the differences of the
parametrized post-Newtonian parameter $\gamma$ values as low as
$\left[\gamma(1.23\; \rm GHz)-\gamma(1.45\; \rm GHz)\right]<4.36\times10^{-9}$.
This provides the most stringent limit up to date on the EEP through the
relative differential variations of the $\gamma$ parameter at radio energies,
improving by 1 to 2 orders of magnitude the previous results at other energies
based on supernova 1987A and GRBs.
\end{abstract}

\maketitle

\section{Introduction}

The Einstein Equivalence Principle (EEP) is an important foundation of general relativity
and many other metric theories of gravity. At the post-Newtonian level, the accuracy of the EEP
can be tested through the numerical values of the parametrized post-Newtonian (PPN) parameters,
such as the parameter $\gamma$ (see, e.g., Refs. \cite{Will2006,Gao2015}). Specifically,
the EEP accuracy can  be constrained by comparing the $\gamma$ values for different kinds of
particles, or for the same kind of particle with different energies, since all gravity theories
satisfying the EEP predict $\gamma_{1}=\gamma_{2}\equiv\gamma$, where the subscripts denote
two different test particles.

There are a few precise tests of the EEP using constraints on the differences of the $\gamma$
values of different tested particles. Among the most famous are the measurements of the time
delay of the photons and neutrinos radiated from supernova 1987A in the Large Magellanic Cloud
\cite{Krauss1988,Longo1988}.
Recently it was shown \cite{Gao2015} that the EEP can also be tested using the time delay
of photons with different energies arising in cosmic transients, such as gamma-ray bursts
(GRBs) \cite{footnote}.
With the assumption that the time delays are mainly due to the Milky Way's gravity,
it was found \cite{Gao2015} that the value of $\gamma$ is identical for photons over an energy range
between eV and MeV and between MeV and GeV to within  approximately $10^{-7}$,
which represents an improvement of at least one order of magnitude compared with
the previous limits.

Recently a new type of millisecond radio burst transients, named fast radio bursts (FRBs),
has attracted wide attention \cite{Lorimer2007,Thornton2013}. Following the
first report of an FRB by Ref. \cite{Lorimer2007}, a number of further FRBs have been reported,
with a present total of over ten cases \cite{Thornton2013,Keane2012}.
Most of these bursts are located at high galactic latitudes and have anomalously large
dispersion measures (DM). The observed event rate is predicted to be $\sim 10^{-3}$
$\rm gal^{-1}$ $\rm yr^{-1}$. Moreover, the higher frequency components of an FRB arrive
earlier than their low frequency counterparts, the arrival time delay at a given frequency
$\nu$ following a $\nu^{-2}$ law \cite{Thornton2013,Karastergiou2015}. Based on these typical characteristics,
it has been suggested that these sources may originate at cosmological distances,
corresponding to redshifts $z$ of 0.5 to 1. If so, the isotropic total energy released in
one FRB is inferred to be $\sim10^{38-40}$ erg, and the peak radio luminosity is estimated
to be $\sim10^{42-43}$ erg $\rm s^{-1}$ \cite{Thornton2013}. However, because there are
no clear-cut electromagnetic counterparts at other wavelengths detected, and no
progenitors have been identified, the physical nature of FRBs remains under debate.
Many possible explanations for FRBs have been proposed, including magnetar flares
\cite{Popov2013}, neutron star mergers \cite{Totani2013}, white dwarf mergers
\cite{Kashiyama2013}, collapsing super-massive neutron stars \cite{Falcke2014,Zhang2014},
companions of extragalactic pulsars \cite{Mottez2014}, asteroid
collisions with neutron stars \cite{Geng2015}, quark nova \cite{Shand2015},
and dark matter-induced collapse of neutron stars \cite{Fuller2015}. All of these models
considered FRBs as extragalactic burst sources. It is worth noting that some other models,
which suggested a Galactic (i.e., Milky Way) origin for FRBs, have also been proposed,
e.g., galactic flare stars \cite{Loeb2014} and atmospheric phenomenon \cite{Kulkarni2014}.

If FRBs are indeed confirmed to be cosmological origin, they will be a useful cosmic probe.
For example, they can be used to determine the baryon mass density of the Universe
\cite{Deng2014}, and they could also be used to constrain cosmological parameters and
the dark energy \cite{Gao2014,Zhou2014}. Here we propose that cosmic originated
FRBs are also good candidates for constraining the EEP, which could further expand the scope
of the tested EEP energy range out to the radio band with high accuracy.

Compared to the prospects of GRBs in constraining the EEP, the FRBs have two advantages.
Firstly, the simple sharp feature of the FRB signal allow us to easily derive the observed
time delay between different frequencies, and their time lags are usually shorter than
that of most GRBs. The value of the time delay plays an important role in constraining the EEP,
a smaller time delay leading to better constraints on the EEP. Secondly, although GRBs are
multi-wavelength transients, it is hard to measure in them the arrival time lag in the radio
band.  A more promising way to further extend the EEP tested energy range down to the radio
band is with the help of FRBs.
In addition, if a fraction of the FRBs originate from the delayed collapse of
supermassive millisecond magnetars to black holes \cite{Falcke2014}, it has been suggested
that these sources could be associate with some GRBs \cite{Zhang2014}. In this case, the
GRB location provides information additional to that of the FRB time delay.
In this paper, we extend the work of Ref. \cite{Gao2015} by presenting stronger constraints
on the EEP using FRBs.

\section{Method of testing the EEP}

For a cosmic transient source, the observed time delay between two different energy bands
should include five terms \cite{Gao2015}:
\begin{equation}
\Delta t_{\rm obs}=\Delta t_{\rm int}+\Delta t_{\rm LIV}+\Delta t_{\rm spe}+\Delta t_{\rm DM}+\Delta t_{\rm gra}\;.
\end{equation}
$\Delta t_{\rm int}$ is the intrinsic (astrophysical) time delay between two test photons,
$\Delta t_{\rm LIV}$ is the time delay caused by the effect of Lorentz invariance violation,
and $\Delta t_{\rm spe}$ represents the potential time delay due to special-relativistic effects
in the case where the photons have a rest mass which is non-zero.
$\Delta t_{\rm DM}$ is the time delay contribution from the dispersion by the
line-of-sight free electron content, which is non-negligible especially for low energy photons,
such as the radio signals considered.
$\Delta t_{\rm gra}$ corresponds to the difference in arrival time of two photons of energy $E_1$
and $E_2$, caused by the gravitational potential $U(r)$ integrated from the emission site to the
Earth, which reads
\begin{equation}
\Delta t_{\rm gra}=\frac{\gamma_{\rm 1}-\gamma_{\rm 2}}{c^3}\int_{r_o}^{r_e}~U(r)dr\;,
\end{equation}
where $r_{e}$ and $r_{o}$ are locations of source and observation, $\gamma$ is the PPN parameter.
For the purposes of this work, both $\Delta t_{\rm LIV}$ and $\Delta t_{\rm spe}$ are negligible,
we thus ignore them in our analysis (see Ref. \cite{Gao2015}, for more explanations).
Leaving out the negligible components, and assuming that $\Delta t_{\rm int}>0$, we have
\begin{eqnarray}
\Delta t_{\rm obs}-\Delta t_{\rm DM}>\frac{\gamma_{\rm 1}-\gamma_{\rm 2}}{c^3}\int_{r_o}^{r_e}~U(r)dr\;.
\label{eq:deltatnew}
\end{eqnarray}

For a cosmic transient, in principle, $U(r)$ has contributions from the gravitational potential
of the Milky Way $U_{\rm MW}(r)$, the intergalactic potential $U_{\rm IG}(r)$ between the
transient host galaxy and the Milky Way, and the potential of the host galaxy $U_{\rm host}(r)$.
The potential models for $U_{\rm IG}(r)$ and $U_{\rm host}(r)$ are poorly known, but it is
very likely that the effect of these two terms is much larger than if we simply assumed that
the potential is just $U_{\rm MW}(r)$ extended to the distance of the transient. Adopting the
Keplerian potential for our galaxy, we have
\begin{equation}
\gamma_{1}-\gamma_{2}<\left(\Delta t_{\rm obs}-\Delta t_{\rm DM}\right)\left(\frac{GM_{\rm MW}}{c^{3}}\right)^{-1}\ln^{-1}\left(\frac{d}{b}\right) \;,
\label{eq:gammadiff}
\end{equation}
where $M_{\rm MW}\simeq6\times10^{11}M_{\odot}$ is the Milky Way mass \cite{McMillan2011}, $d$ is the distance from the source to the Earth, and $b$ is the impact parameter of the light rays relative to the Milky Way center \cite{footnote2}.

\section{Tests of the EEP using FRBs}

As mentioned above, if FRBs are proven to be of cosmological origin and if their distances
can in the future be accurately measured, FRBs will be a new powerful tool for obtaining
EEP constraints, while extending the tested energy range down to the radio band.

The single-dish telescope used to detect all but one of the currently known FRBs localizes
these sources to about 0.25 degrees \cite{Thornton2013}, and hence electromagnetic counterparts,
if any, such as GRBs, may be critical to determine distances and thereafter analyze source
energetics and estimate event rates \cite{Zhang2014, Yi2014}.

So far, there are two main methods to estimate the distances of FRBs.  The first method
(Method 1) is to directly estimate the redshifts of the FRBs through their DM values (e.g., Ref.
\cite{Thornton2013}). In this method, some delicate assumptions are adopted, which can result
in a large uncertainty for the source distance. For example, to estimate the DM value,
$\Delta t_{\rm obs}\approx\Delta t_{\rm DM}$ is assumed. However, if the main contribution to
$\Delta t_{\rm obs}$ is from $\Delta t_{\rm gra}$ instead of $\Delta t_{\rm DM}$, this method
would severely overestimate the source distance. On the other hand, to connect the cosmic
distances of FRBs with their DM values, the contributions of the DM dispersion from the source
environment and host galaxy are assumed to be negligible compared to the intergalactic medium
(IGM) component. If this is not the case, the source distance would also be overestimated.
Finally, to calculate the distance with the IGM DM value, one needs to assume that the average
IGM DM value in all directions for a given \emph{z} is well defined by known cosmological
parameters \cite{Deng2014,Gao2014}. The uncertainty of the relevant cosmological parameters
could cause either under- or over- estimation of the source distance.

The second method (Method 2) is to make use of the redshifts of possible FRB/GRB
association systems. In this method, the FRB distance can  be precisely estimated when the
associated GRB has a direct redshift measurement (from the GRB afterglows or the GRB host
galaxies ). If not, some empirical luminosity relations, such as the Amati relation
\cite{Amati2002}, may be used to give a rough range of pseudo-redshifts for the FRB/GRB system
(e.g., Ref. \cite{Deng2014}).

Here we take the following two examples, FRB 110220 (one of the brightest bursts, with the
clearest frequency-dependent delays) and two potential FRB/GRB association systems
(FRB/GRB 101011A and FRB/GRB 100704A), and use these to constrain the EEP.

\subsection{FRB 110220}
In a recent search for pulsars, four FRBs have been discovered by the 64-m Parkes multibeam
radio telescope \cite{Thornton2013}. FRB 110220 is one of the brightest bursts, with a flux
$\sim2.5$ Jy (at 1.5 GHz), and it was detected at $T_{0}=01:55:48.957$ UT, February 20th,
2011, with coordinates (J2000) $\rm R.A.=22^{h}34^{m}$, $\rm Dec.=-12^{\circ}24^{'}$.
From the frequency-dependent delay of FRB 110220 (see Figure~2 of \cite{Thornton2013}),
one can easily identify the arrival time delay $\Delta{t}_{\rm obs}\sim1$ s for photons ranging
in frequency from about 1.5 GHz to 1.2 GHz. With Method 1, this source is inferred to be at a
redshift $z_{\rm infer}\sim0.81$.

With the above information, from Equation~(\ref{eq:gammadiff}) a severe limit on the EEP is
\begin{equation}
\left[\gamma(1.2\; \rm GHz)-\gamma(1.5\; \rm GHz)\right]<2.52\times10^{-8}
\end{equation}
for FRB 110220, which is almost $10^{2}$ times tighter than the constrains from supernova 1987A,
and is as good as the results on GRBs from Ref. \cite{Gao2015}.
Note that this is a relatively conservative upper limit: we assume $\Delta t_{\rm obs}
\approx\Delta t_{\rm DM}$ when estimating the redshift (which is inferred from the characteristic
observational feature of FRBs, i.e., that the arrival time delay at a given frequency $\nu$ follows
a $\nu^{-2}$ law), but we take $\Delta t_{\rm obs} \gg \Delta t_{\rm DM}$ for a stringent limit
on the EEP. For completeness, we also tested two more cases by assuming
$\Delta t_{\rm DM}=0.001\Delta t_{\rm obs}$ and $\Delta t_{\rm DM}=0.999\Delta t_{\rm obs}$.
As shown in Figure 1, much more severe constraints would be achieved ($\sim10^{-11}$) if the
effects of the dispersion process represent $99.9\%$ of $\Delta t_{\rm obs}$. Even if they
represent only 0.1\% of $\Delta t_{\rm obs}$, the limits we derived here are still about
$\sim10^{-8}$.

To account for the source distance uncertainty, we also test the results by varying the
source distance from 1 Mpc (the edge distance of the Local Group) to 3$z_{\rm infer}$. As shown
in Figure 1, we find that even if the distance estimation for FRBs have large uncertainties,
the implications of the FRB tests of the EEP are not greatly affected, e.g., the constraint
results vary within one order of magnitude.

\subsection{FRB/GRB systems}
Recently, two possible associations of FRBs with
GRB 101011A and GRB 100704A have been proposed by Ref. \cite{Bannister2012}. If such FRB/GRB
association systems are commonly discovered, the combination of redshifts measured from GRBs
and DMs measured from FRBs not only opens a new window to study cosmology \cite{Deng2014,Gao2014},
but also makes them an interesting tool for EEP constraints (more on this below).

For these two FRB/GRB association systems, one can in principle use their location information
from the GRB observations. GRB 101011A was detected and localized by $Swift$/BAT at
$T_{0}=16:58:35$ UT, October 11th, 2010, with coordinates (J2000) $\rm R.A.=03^{h}13^{m}12^{s}$,
$\rm Dec.=-65^{\circ}59^{'}08^{''}$ \cite{Cannizzo2010}. At $T_{0}=03:35:06.10$ UT on
04 July 2010, the $Fermi$ Gamma-Ray Burst Monitor triggered and located GRB 100704A
\cite{McBreen2010}, which was also detected by $Swift$/BAT with coordinates (J2000)
$\rm R.A.=08^{h}54^{m}33^{s}$, $\rm Dec.=-24^{\circ}12^{'}55^{''}$ \cite{Grupe2010}.
Unfortunately, neither of  these two GRBs had a redshift measurement. Ref. \cite{Deng2014}
used the Amati relation to estimate the redshifts of these two FRB/GRB association systems:
$z\geq0.246$ for GRB 101011A and $z\geq0.166$ for GRB 100704A.
Ref. \cite{Bannister2012} observed these two systems over a 220 MHz bandwidth, with a highest
frequency of $\nu_h=1.45$ GHz and lowest frequency of $\nu_l=1.23$ GHz. The delay times between
these two frequencies are $\Delta t_{\rm obs}=0.149$ s for FRB/GRB 100704A and
$\Delta t_{\rm obs}=0.438$ s for FRB/GRB 101011A.

With the inferred redshifts for the two FRB/GRB systems (here we adopt
$z_{\rm infer}=0.246$ for GRB 101011A and $z_{\rm infer}=0.166$ for GRB 100704A)
and assuming $\Delta t_{\rm obs} \gg \Delta t_{\rm DM}$, a stringent limit on the EEP
from Equation~(\ref{eq:gammadiff}) is
\begin{equation}
\left[\gamma(1.23\; \rm GHz)-\gamma(1.45\; \rm GHz)\right]<1.24\times10^{-8}
\end{equation}
for GRB 101011A, and
\begin{equation}
\left[\gamma(1.23\; \rm GHz)-\gamma(1.45\; \rm GHz)\right]<4.36\times10^{-9}
\end{equation}
for GRB 100704A.

We also tested these two results by assuming $\Delta t_{\rm DM}=0.001\Delta t_{\rm obs}$
and $\Delta t_{\rm DM}=0.999\Delta t_{\rm obs}$, and by varying the source distance from
the edge distance of the Local Group to 3$z_{\rm infer}$. We find that considering these
uncertainties, the limits on the EEP stay at the level of $\sim10^{-8}$, and much
more severe constraints could be achieved if it turns out that the effects of dispersion process
dominate the observed time delay.
Note also that the EEP test energy range could in principle be extended further by using
the measured time difference between the GRB itself and the FRB, although this would involve
additional astrophysical uncertainties, which we do not consider here.

\section{Summary and discussion}

The accuracy of the EEP can be characterized by constraints on the differences in PPN
parameters, such as the parameter $\gamma$, for different kinds of particles, or for
the same kind of particle with different energies. Following the method of Ref. \cite{Gao2015},
we prove that FRBs, if cosmological, can be used to test the accuracy of the EEP,
leading to 1-2 orders of magnitudes stricter limits than previously.

From the arrival time delay $\Delta{t}\sim1$ s for photons ranging in frequency from about
1.2 GHz to 1.5 GHz, assuming that the observed time delay is caused mainly by the
gravitational potential of the Milky Way and adopting the inferred redshift either based
on DM measurement or based on associated GRB observations, we obtain stringent limits on
the differences of the $\gamma$ values for three FRB cases:
$\left[\gamma(1.2\; \rm GHz)-\gamma(1.5\; \rm GHz)\right]<2.52\times10^{-8}$ for FRB 110220,
$\left[\gamma(1.23\; \rm GHz)-\gamma(1.45\; \rm GHz)\right]<1.24\times10^{-8}$ for FRB/GRB 101011A
and $\left[\gamma(1.23\; \rm GHz)-\gamma(1.45\; \rm GHz)\right]<4.36\times10^{-9}$ for
FRB/GRB 100704A.

\begin{figure}[h]
\epsfig{figure=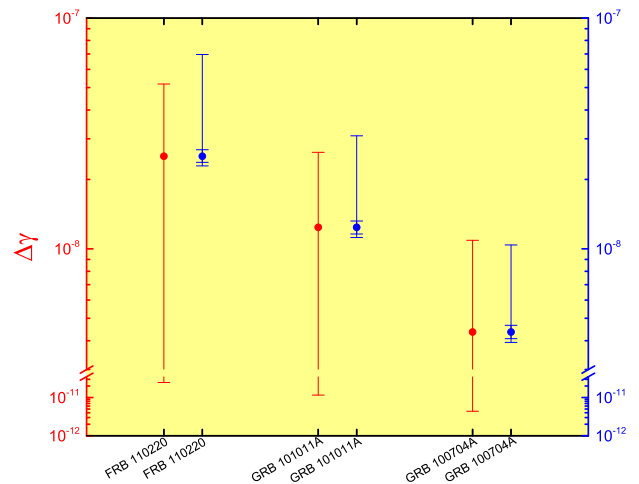,width=3.5truein}
\vskip-0.2in
\caption{Limits on the differences of the $\gamma$ values for three FRB observations.
The dots represent the conservative upper limits on $\Delta\gamma$ by taking the inferred
redshifts from Method 1 or 2, and assuming $\Delta t_{\rm obs} \gg \Delta t_{\rm DM}$.
The red lines show the test results from considering different $\Delta t_{\rm DM}$ contributions.
From top to bottom, the red error bars correspond to the case of $\Delta t_{\rm DM}=0.001\Delta t$
and $\Delta t_{\rm DM}=0.999\Delta t$, respectively.  The constraints on $\Delta\gamma$ from
varying the values of redshifts are presented as the blue lines.  From top to bottom, the blue
error bars correspond to the cases of 1 Mpc, $0.5z$, $2z$, and $3z$, respectively.
}
\end{figure}

Previously, by analysing the photons and neutrinos from supernova 1987A,
Ref. \cite{Longo1988} set a severe limit on $\gamma$ differences of $0.34\%$ for
optical photons (eV) and neutrinos (MeV), and a more precise limit of $1.6\times10^{-6}$
for two neutrinos with different energies (7.5 -- 40 MeV).
Very recently, Ref. \cite{Gao2015} proposed that this EEP parameter can also be tested with
the time delays between correlated photons from GRBs. Compared with previous limits,
their constraint on the accuracy of the EEP of $\sim10^{-7}$ represented an improvement
of at least one order of magnitude, extending also the tested energy range to the
eV-MeV and MeV-GeV range.
In the present paper, using the sharp features of the FRB radio signals, we have
extended the energy range over which the EEP is tested to the radio band at comparable
or higher levels of accuracy, and we have obtained the most stringent limit to date on
the EEP, namely $\sim10^{-8}$, which represents an improvement of one to two orders of
magnitude over the results previously obtained at other energies using supernova 1987A
and GRBs.

Note that this is a conservative upper limit: the inclusion of contributions from the
neglected potentials in Equation~(1) could only make these limits even more stringent.
On the other hand, if the time delay between different frequencies is mainly
contributed by the dispersion process by the line-of-sight free electron content,
much more severe constraints could be achieved (e.g., $\Delta \gamma \sim10^{-11}$)
if the effects of dispersion process represent $99.9\%$ of $\Delta t_{\rm obs}$.
It is worth noting also that large uncertainties in the source distance would not
significantly affect our conclusions, as long as the FRBs originate outside
of the Local Group.

The impact of the results presented in this work is expected to increase
significantly as more FRBs are observed, and if their origin is more firmly established.
Based on current observations, the FRB event rate is estimated to be as high as
$\sim10^{4}\; \rm sky^{-1}\;day^{-1}$ \cite{Thornton2013}. The current low detection rate
of FRBs may be due to the lack of either the necessary high time resolution or a wide enough
field of view in the current telescopes. Fortunately, the upcoming radio transient surveys
such as the Square Kilometer Array, with much wider field of view and higher sensitivity,
will be able to discover and precisely localize more and more FRBs. With more abundant
observational information in the future, we will have a better understanding of the physical
nature of FRBs, and we will also know better how to use them as a probe of the cosmic web,
the host galaxies, the intergalactic medium and to test fundamental physics,
as discussed in this Letter.

\vskip 0.1in
\noindent{\bf Acknowledgements:}
We are grateful to the anonymous referees for insightful comments, and to K. W. Bannister
for helpful communications.
This work is partially supported by the National Basic Research Program (``973'' Program)
of China (Grants 2014CB845800 and 2013CB834900), the National Natural Science
Foundation of China (grants No. 11322328, No. 11433009, and No. 11543005), the One-Hundred-Talents Program,
the Youth Innovation Promotion Association (2011231), the Strategic Priority Research Program
``The Emergence of Cosmological Structures'' (Grant No. XDB09000000) of the Chinese Academy
of Sciences, and NASA NNX13AH50G.


\begin{thebibliography}{99}
\bibitem{footnote2} Equation (4) is only valid when the impact parameter $b$ is large, say larger than the radius of Einstein ring, otherwise the right hand side of the equation approaches non-convergence.
\end{thebibliography}
\end{document}